\documentclass[12pt]{article}
\usepackage{graphicx}
\usepackage{cite}
\def\be{\begin{equation}}
\def\ee{\end{equation}}
\def\bea{\begin{eqnarray}}
\def\eea{\end{eqnarray}}
\def\ba{\begin{array}}
\def\ea{\end{array}}

\def\nn{\nonumber}

\def\Ga{\Gamma}

\def\nn{\nonumber}

\def\nin{\noindent}

\begin{document}
\hfill\parbox{4cm}{\normalsize CERN-TH/2003-122}
\unitlength1cm

\begin{center}

{\LARGE {\bf Effective action and the quantum equation of motion}}\\

\vspace*{0.5cm}
Vincenzo Branchina\footnote{Vincenzo.Branchina@cern.ch}

{\it Theory Division, CERN, CH-1211  Geneva 23,  Switzerland}\\

and 

{\it IReS Theory Group, ULP and CNRS, 23 rue du Loess, 67037 Strasbourg, France}\\
\vspace*{0.1cm}
Hugo Faivre\footnote{Hugo.Faivre@ires.in2p3.fr}

{\it IReS Theory Group, ULP and CNRS, 23 rue du Loess, 67037 Strasbourg, France}\\

\vspace*{0.1cm}
Dario Zappal\`a\footnote{Dario.Zappala@ct.infn.it}\\

{\it INFN, Sezione di Catania, Via S. Sofia 64, I-95123, Catania, Italy}\\

and 

{\it Dip. di Fisica, Universit\`a di Catania, 
Via S. Sofia 64, I-95123, Catania, Italy}\\

\vspace*{0.4cm}

{\LARGE Abstract}\\

\end{center}

We carefully analyse the use of the effective action in dynamical 
problems, in particular the conditions under which 
the equation $\frac{\delta \Ga} {\delta \phi}=0$ can be used 
as a quantum equation of motion,
and illustrate in detail the crucial  relation between the
asymptotic states involved in the definition of $\Ga$ and the initial 
state of the system. Also,
 by considering the quantum mechanical example of a
double-well potential, where we can get exact results for the time
evolution of the system, we show that an approximation to 
the effective potential in the quantum equation of motion that correctly 
describes the dynamical evolution of the system is obtained with the help 
of the wilsonian RG equation (already at the lowest order of the derivative 
expansion), while the commonly used one-loop effective potential fails to 
reproduce the exact results.

\vskip 10pt
Pacs 11.10.Gh, 98.80.Cq, 03.65.Db 

\newpage

\section{Introduction}

The effective action $\Gamma[\phi]$, the generating functional of the
one-particle irreducible (1PI) vertex functions, is a very useful tool 
in quantum field theories. It is widely used in the analysis of their 
vacuum structure\cite{jona} (for more recent reviews see e.g. \cite{rivers,lowell}), 
and its symmetries are often exploited 
in order to establish their renormalization properties\cite{zin}. For 
vanishing external sources, it satisfies the equation 

\be
\label{qeq}
\frac{\delta \Ga [\phi]} {\delta \phi(\vec x, t)} =0,
\ee

\nin
sometimes referred to as the quantum equation of motion, the quantum 
counterpart of the classical equation 
$\frac{\delta S} {\delta \phi} =0$.

In fact, at the lowest order of the semiclassical expansion, 
$\Gamma[\phi]$ coincides with the classical action $S[\phi]$, and this
property, when combined with Eq. (\ref{qeq}), naively suggests that
$\Ga[\phi]$ should be regarded as the quantum action of the system, and 
Eq.(\ref{qeq}) as the corresponding quantum dynamical equation. This 
interpretation, however, is correct only under certain conditions.
Moreover, even when this interpretation applies, 
Eq.(\ref{qeq})  in general cannot be straightforwardly  
related to any  semiclassical expansion. 

The dynamical content of Eq.(\ref{qeq}) was already  
briefly discussed in \cite{noiprl}, where a solid theoretical background 
to the work presented in \cite{lope} was provided along with the
correction of some mistakes and misunderstandings. One of the purposes of 
the present paper is to present a more detailed description of the approach
followed in \cite{noiprl}, with the aim of clarifying the limits 
of applicability of this procedure. 

As it is well known, an ansatz for $\Gamma[\phi]$ which is well 
suited for the applications is given by the gradient expansion,  
an expansion in powers of the field derivatives :

\bea\label{grad}
\Ga[\phi] = \int {\rm d}^4x \bigg(&-&V_{eff}(\phi)+
\frac{Z(\phi)}{2}\partial_\mu\phi
\partial^\mu\phi + {Y(\phi)}(\partial_\mu\phi\partial^\mu\phi)^2 \nn \\
&+&W(\phi) (\partial_\mu\partial_\mu\phi)^2 + \cdots \bigg)\, . 
\eea

The lowest order approximation, the so called Local Potential 
Approximation (LPA), is obtained from Eq.(\ref{grad}) once we neglect
the higher order derivatives and keep the wave function renormalization 
factor constant, i.e. field independent (without loss of generality we 
can choose $Z=1$) : 

\be\label{lpa}
\Ga[\phi]=\int {\rm d}^4x \biggl(-V_{eff}(\phi(x))+\frac12\partial_\mu\phi(x)
\partial^\mu\phi(x)\biggr) \, .
\ee

\nin
Within the LPA, Eq.(\ref{qeq}) becomes:

\be\label{eqlpa}
\partial^\mu\partial_\mu\phi = -\frac{\partial V_{eff}(\phi)}
{\partial\phi} \, ,
\ee

\nin
i.e. it takes the same form of the classical equation of motion 
where the classical potential is replaced by the effective potential. 

This equation has been widely used in the past
as the quantum equation of motion to describe the 
time evolution of the scalar field expectation value\cite{brande}. 
More recently, however, some of the limitations in the use of 
Eq.~(\ref{eqlpa}), and more generally of Eq.~(\ref{qeq}), as a dynamical 
equation have been noted\cite{devega} and a different formalism to deal 
with dynamical problems has been developed\cite{cooper, pi, devega}. 

In the present paper we complement these previous studies and push 
the investigation of Eq.~(\ref{eqlpa}) a step further. We carefully 
analyse the limitations in the 
application of this equation to dynamical problems and show that 
there are physically relevant cases where 
it can be appropriately used to describe the evolution 
of expectation values. It will turn out that two points, 
overlooked in the past, play a crucial role: 
the correct consideration of the boundary conditions encoded 
in the definition of the effective action and the use of 
a non perturbative approximation for the effective potential 
in Eq.~(\ref{eqlpa}).
  
Let us proceed now to our systematic analysis of the subtle points 
related to the use of Eq.~(\ref{qeq}) as a dynamical equation. As it will
be clear in a moment, it is essential to specify:
\par\noindent
(i) the framework in which Eq.~(\ref{qeq}) is derived, to avoid confusion on 
the physical meaning of  $ \phi$;
\par\noindent
(ii) the choice of the boundary  conditions (consistent with point (i))
associated to the differential equation (\ref{qeq});
\par\noindent
(iii) the approximation in which $\Gamma[\phi]$ is computed, that has obviously 
to be well suited for the particular problem considered, in order to get 
physically meaningful results.

Concerning point (i), one has to be careful in considering  
$ \phi$ as the field expectation value because in general this is not true.
The effective action $\Ga[\phi]$ is the Legendre transform of $W[J]$, 
the generator  of the connected Green functions, defined by:
$e^{\frac{i}{\hbar}W[J]} = Z[J]= \langle0, t=+\infty|0, t=-\infty\rangle_J$.
In the Schr\"{o}dinger picture, the vacuum persistence amplitude, 
$Z[J]$, can be explicitly written as:

\be\label{zsch}
Z[J]= \langle0|\hat U(+\infty, -\infty)|0\rangle = 
\langle0|T\Bigl(e^{-\frac{i}{\hbar}
\int_{-\infty}^{+\infty}{\rm d}t(\hat H - J(t)\hat\Phi(\vec x))}\Bigr)|0\rangle\, ,
\ee
\nin
where $\hat U(+\infty, -\infty)$ is the time 
evolution operator in the presence of the source $J$, and $|0\rangle$ is 
the ground state of $\hat H$, the Hamiltonian of the system. 
\nin
From Eq. (\ref{zsch}), and from the definition of the classical field, 
$\phi(\vec x, t)=\frac{\delta W[J]}{\delta J(\vec x, t)}$, we immediately 
see that: 

\be\label{cla}
\phi(\vec x, t) = \frac{\langle -,t| \hat\Phi(\vec x)| +,t\rangle  }
{\langle -,t| +,t\rangle},
\ee

\nin
where 
\be\label{states}
\langle -,t|=\langle0|\hat U(+\infty, t) ~~~~~~~~~~~~~~~~~~ {\rm and} 
~~~~~~~~~~~~~~~~~~ | +,t\rangle= \hat U(t, -\infty)|0\rangle\, . 
\ee

\nin
Clearly  $|+,t\rangle$ and $|-,t\rangle$ are (in general) 
different states. As a consequence, $\phi(\vec x, t)$ is not (in general)
a diagonal matrix element of the quantum field $\hat \Phi(\vec x)$.  
Moreover it may be complex-valued. This is not surprising 
as $\Ga$ generates the 1PI vertex functions that are in general complex 
quantities satisfying causal (Feynman) boundary conditions. Only for particular 
choices of the external source $J$ can we have $|+,t\rangle =
|-,t\rangle$; under these conditions, Eq. (\ref{qeq}) describes the 
dynamical evolution of the field expectation value of $\hat\Phi$. 

Effective actions defined according to Eq. (\ref{zsch}) are called 
Schwinger--De Witt
(or ``in--out") effective actions\cite{dewitt}. 
They incorporate boundary conditions that are appropriate to  
problems where the transition from asymptotic states in the past 
to asymptotic states in the future is considered
(as is the case of a scattering process).
In passing we note that more general definitions of ``in--out" 
effective actions can be obtained if the $|0, t=-\infty\rangle$ and  
$|0, t=+\infty\rangle$ asymptotic states are replaced by 
more general $|{\rm in}\rangle$ and $|{\rm out}\rangle$ states, 
i.e. by more general asymptotic boundary conditions. 

A functional formalism that is in general appropriate to dynamical 
problems is the so called ``in--in" or closed-time-path 
formalism\cite{sch,bak,cal}, where one can construct ``in--in" 
effective actions that generate the dynamical evolution 
of expectation values. This formalism is the one most
largely used nowdays in the applications to inflationary cosmology.

In this paper, however, we shall not deal with this latter approach 
and its most recent developments. Our aim is rather to show that, 
under certain conditions, we can still define an equation of motion 
in the in-out formalism. 

In view of the fact that Eq.~(\ref{qeq})
is sometimes used as the quantum counterpart of the classical 
equation of motion, this problem is certainly worth to study. 

Even in those cases where this formalism can be properly employed,
however, we still have to face the  problem mentioned in point (ii). 
Equation (\ref{zsch}) contains two asymptotic conditions 
at the initial ($t=-\infty$) and final ($t=+\infty$) times. These 
conditions are encoded in the definition of the effective action 
itself, as well as in the definition of the classical field 
(the argument of $\Ga$). It is then clear that we cannot 
freely choose certain boundary conditions for the field and its derivatives 
at a given initial time $t=t_0$, together with a ``physically convenient" form for 
the (functional) wave packet of the system at the same time, and then  
evolve the expectation value of the quantum field according to Eq. (\ref{qeq}). 
In fact arbitrary  initial conditions for $\phi$ and for  the
wave packet in general are not compatible with 
the asymptotic conditions that enter the definition of $\Ga$. 

In the following, we provide arguments showing that, 
by considering a specific set of initial conditions 
for the expectation value of the field 
and of its derivatives, it is possible to find the form 
of the initial (i.e. at $t=t_0$) wave packet that is compatible with 
the asymptotic conditions encoded in the definition of the 
effective action, and whose dynamical evolution is 
governed by Eq. (\ref{qeq}).
Our arguments will be strongly supported by the numerical 
results that we shall obtain for the dynamical evolution of the 
position operator expectation value in the quantum mechanical 
double-well potential problem. In this case we are also able to
consider  
the exact dynamical evolution with the help of the time-dependent 
Schr\"{o}dinger equation, so that we can compare the results obtained 
with Eq. (\ref{qeq}) with the exact ones. As we shall see, when the 
initial wave packet is chosen according to our criterion, we find 
excellent agreement between the two results.

We have found it convenient to investigate  
the question of the determination of the initial wave packet
within the framework of the variational definition of the 
effective action\cite{jac}, that is the generalization to 
the time-dependent case of the well-known variational definition 
of the effective potential\cite{colem}. The effective action 
$\Ga[\phi]$ is obtained through the extremization of a certain 
functional of two (a priori) different states, called $|\psi_+,t\rangle$ 
and $|\psi_-,t\rangle$, under the constraint that 
$\langle\psi_+,t|\hat\Phi(\vec x)|\psi_-,t\rangle = \phi(\vec x,t)$ 
(see next section).

Finally we come to point (iii) and to the problem of the approximations employed 
to compute the effective action and effective potential.
The most simple approximation of $\Ga$ is given by the LPA, i.e. by 
Eq.(\ref{lpa}), and the most straightforward approximation 
of the effective potential $V_{eff}$ 
is given by the one-loop potential $V_{1l}$.
This last approximation however presents a serious drawback. 
The exact effective potential (actually the effective 
action) is a convex function of its argument \cite{sim,curt,rivers}. However, when the 
classical (bare) potential is not convex (these are the physically 
most interesting cases), at any finite order of the loop expansion 
the approximated effective potential does not enjoy this 
fundamental property. 
Alternative non-perturbative methods of computing the effective action and potential,
though, such as lattice simulations \cite{latti}, variational approaches \cite{noivar},
or suitable averages of the perturbative results \cite{wewu},  
provide the proper convex shape. 
In addition to the methods quoted above,
a non-perturbative convex approximation to the effective potential, 
$V_{RG}$, is found within the framework of the wilsonian renormalization 
group (RG) equation\cite{fuku,ring,tet1,ale,tet2}. We have computed the effective 
potential with the help of this RG equation and then 
inserted $V_{RG}$ in Eq.~(\ref{eqlpa}). 
The comparison of the results obtained with Eq. (\ref{qeq}) with those
obtained with the help of the time-dependent Schr\"{o}dinger equation
shows that  $V_{RG}$, because of  its non-perturbative features, provides an excellent 
approximation for  the correct ``quantum potential" to be used in 
the ``quantum equation of motion". 
For completeness we will also check the inadequacy of
$V_{1l}$ to describe non-perturbative regimes with our equations of motion.

The plan of the paper is as follows. In Section 2 we briefly 
illustrate the basic formalism employed in the following. 
In Section 3 the central argument concerning the validity of 
Eq. (\ref{qeq}) is discussed, while its application to the  
harmonic oscillator and to the double-well potential are respectively addressed  
in Section 4 and 5. The former example is treated analytically whereas the latter is 
solved by a numerical  analysis. Section 6 contains the summary and
outlook.

\section{Variational definitions of $\Gamma [\phi]$ 
and $V_{{eff}}(\phi)$ }

To set up the tools of the following analysis, we briefly review 
in this section the variational principles that lead to the 
definitions of the effective action and the effective potential,
referring to\cite{jac} and \cite{colem} for details. 
According to \cite{jac} the effective action is the stationary, time-integrated 
matrix element of ~$i \partial_t -\hat H$ 
\begin{equation} 
\label{pvea}
\Gamma[\phi]=\int^{+\infty}_{-\infty} {\rm d}t  
\langle \psi_- ,t| \left (i \partial_t -\hat H \right ) | \psi_+
,t\rangle ,
\end{equation}
where the right-hand side is stationary when the two time-dependent states 
$| \psi_\pm ,t\rangle$ are varied arbitrarily and independently, 
but with the two constraints

\begin{equation} \label{c1}
\langle \psi_- ,t| \hat \Phi({\vec x}) | \psi_+ ,t\rangle =\phi({\vec x},t)
\end{equation}
and 
\begin{equation} \label{c2}
\langle \psi_- ,t | \psi_+ ,t\rangle =1,
\end{equation}
and with  the asymptotic boundary conditions
\be\label{condi}
\lim_{t\to \mp \infty} | \psi_\pm ,t\rangle=|0 \rangle,
\ee
where $|0 \rangle$ is the ground state of $\hat H$.

This variation, together with the 
constraints (\ref{c1}) and  (\ref{c2}), is translated into the
equations:
\begin{equation} \label{eqstd1}
\left (i \partial_t -\hat H +\int {\rm d}^3{\vec x } ~J({\vec x},t ) \hat \Phi({\vec x})\right )|\psi_+ ,t\rangle=
w(t)|\psi_+ ,t\rangle
\end{equation}

\begin{equation} \label{eqstd2}
\left (i \partial_t -\hat H +\int {\rm d}^3{\vec x} ~J({\vec x},t) \hat \Phi({\vec x})\right )|\psi_- ,t\rangle=
w^*(t)|\psi_- ,t\rangle,
\end{equation}
where $J({\vec x},t)$ and $w(t)$ are the Lagrange multipliers that
implement the two constraints  (\ref{c1}) and  (\ref{c2}).
The relation between the couple of states $|\psi_\pm ,t\rangle$ and $|\pm ,t\rangle$, 
introduced in 
Eqs. (\ref{states}), is shown in \cite{jac}. They are related by a phase factor 
given by the time integral of the Lagrange multiplier   $w(t)$.

If we limit ourselves to considering constant (in space and time)  
field configurations, the effective action is reduced 
(see Eq. (\ref{lpa})) to the effective potential, the generator 
of the 1PI graphs with vanishing external momenta:
\begin{equation} 
\label{dep}
\left. \Gamma[\phi]\right |_{\phi=const.}=-V_{eff}(\phi) \int {\rm d^4}x.  
\end{equation}

As is well known, $V_{eff}(\phi)$ can be obtained by minimizing 
the expectation value of the Hamiltonian among the normalized 
time-independent states which have a field expectation value equal to 
$\phi$ \cite{colem}, i.e.:

\begin{equation} 
\label{pvep}
V_{eff}(\phi) ={\rm min}_{\psi} \langle \psi| \hat H |\psi \rangle ,
\end{equation}
with the states $ |\psi \rangle$ subject to 
the constraints:
\begin{equation} \label{cs1}
\langle \psi | \hat \Phi({\vec x}) | \psi \rangle =\phi
\end{equation}
and 
\begin{equation} \label{cs2}
\langle \psi | \psi\rangle =1.
\end{equation}
The constrained minimum condition in Eq. (\ref{pvep}) generates the
time-independent 
Schr\"{o}dinger equation for a modified Hamiltonian:  
\begin{equation} \label{eqsti}
\left ( \hat H - J \int {\rm d}^3{\vec x} ~ \hat \Phi({\vec x})\right )|\psi\rangle=
 E |\psi \rangle ,
\end{equation}
where again $J$ and $E$ are the Lagrange multipliers associated to the 
two constraints (\ref{cs1}) and (\ref{cs2}).
As is clear from  Eq. (\ref{pvep}), the ground state(s) of $\hat H$,
which is (are) obtained by  solving Eq. (\ref{eqsti}) with $J=0$,
is (are) associated to the minimum (minima) of $V_{eff}(\phi)$.

Except for very few cases, the exact form of the effective action 
cannot be determined, and we have to resort to some  
approximation. A typical ansatz for $\Ga$, very appropriate for 
our following considerations, is the gradient expansion, 
an expansion in terms of the field derivatives. 
Suitable approximations are obtained by considering truncations
to a given order. 
Although in the applications we shall limit ourselves to considering the 
LPA (see sections 4 and 5), which is the lowest order in this expansion 
and is the approximation typically adopted in dynamical problems,
the considerations that follow do actually apply to any
order of the expansion (see next section).           

\section{The quantum equation of motion}

We have already noted that the classical field $\phi(\vec x,t)$ (the 
argument of $\Ga$) is not in general a diagonal matrix element,
i.e. it is not the expectation value of the quantum field $\hat\Phi$ in 
a given state. However, as we shall see at the end of this section,
under certain conditions the two states $| \psi_+ ,t\rangle$ and $| \psi_- ,t\rangle$ 
of Eq. (\ref{pvea}) coincide: $| \psi_+ ,t\rangle= | \psi_- ,t\rangle =| \psi ,t\rangle$, 
in which case $\phi(\vec x,t)$ is the expectation value of $\hat\Phi(\vec x)$ 
in  $|\psi ,t\rangle$.
For the moment we assume that we are under these conditions.

Let us now consider the differential equation for $\phi(\vec x,t)$,
Eq. (\ref{qeq}),
once the effective action is approximated with a derivative expansion truncated at 
the order $m$, where $2m$ indicates the highest number of field derivatives 
(in the approximation considered in Eq. (\ref{lpa}), it is $m=1$). In
order to get a unique  solution of Eq. (\ref{qeq}) in the time interval $[t_0,t_1]$ 
(where $t_1$ is arbitrarily chosen), we need $2m$ boundary conditions,
which we can fix for instance on the manifold $t=t_0$, with $t_0$ chosen as the 
initial time for the evolution of $\phi$. 

To make contact with the variational principle discussed in Section 2, we have to 
consider the evolution of $\phi(\vec x,t)$ from $t=-\infty$ to
$t=+\infty$, and therefore 
we must provide $\phi(\vec x,t)$ also in the time intervals $]-\infty, t_0]$ and 
$[t_1, +\infty[ $. We can choose the function $\phi(\vec x,t)$ in 
these intervals arbitrarily, provided the asymptotic conditions in Eq.
(\ref{condi}) and the 
proper matching of this function with the unique solution of Eq. (\ref{qeq}) 
in the interval $[t_0,t_1]$ at the times $t_0$ and $t_1$ are taken into account. 
Once the function $\phi(\vec x,t)$ is assigned in the whole range
$]-\infty, +\infty[ $, the constraint on the right-hand side of Eq. (\ref{c1})
is defined and we are therefore able to implement  the variational principle 
and to determine the corresponding source, which we indicate with  
$\overline J(\vec x,t)$, as well as the 
state $| \psi_+ ,t\rangle= | \psi_- ,t\rangle =| \psi ,t\rangle$.

We note in passing that the source $J(\vec x,t)$ can also be  
obtained from its well-known 
relation with the functional  derivative of the effective action 
\be
\label{qeq2}
\frac{\delta \Ga [\phi]} {\delta \phi(\vec x, t)} =J(\vec x, t).
\ee
Equation (\ref{qeq2}) is the generalization of Eq. (\ref{qeq}), when a source, linearly coupled 
to the field, is turned on. Obviously the source  $J(\vec x,t)$ associated to the 
field $\phi(\vec x,t)$ considered above must vanish in the range $[t_0,t_1]$.

So, on the basis of the variational principle, we conclude that there is only one 
state $|\psi, t\rangle$ that is associated to the specific  function  $\phi(\vec x,t)$ 
introduced above, and in particular that at $t=t_0$ the boundary
conditions considered for the time evolution of the field are in fact related to 
$|\psi, t_0\rangle$. This is an important point because, as mentioned in the Introduction, 
if the constraints on the variational principle (see Eqs. (\ref{c1}), 
(\ref{c2}) and (\ref{condi})) are  neglected, one can in general 
find more than one normalized state 
that is compatible with the set of boundary conditions fixed at $t_0$.
However, among these
states, only $|\psi, t\rangle$ is truly related to the effective action 
and therefore to the full evolution of $\phi(\vec x,t)$ in the time 
interval $]-\infty, +\infty[$.

We can now understand what we have already anticipated in the
Introduction concerning the applications. In a 
typical cosmological application the effective action is truncated as in 
Eq.(\ref{lpa}), a convenient form of the wave packet at the initial time $t=t_0$ 
(typically gaussian) is taken, and the initial values for the expectation value of
the scalar field $\phi$ and of its first time derivative are usually
taken to be $\phi={\rm constant}$  and $\dot\phi=0$. On the basis of the previous 
considerations, we easily understand that the time evolution of $\phi$ is not 
always related to the time evolution of the wave packet that has been considered, 
or, in other words, $\phi(\vec x,t)$ need not be the mean value of
$\hat\Phi(\vec x)$ in $|\psi, t\rangle$. 

Our main problem is then the determination of $|\psi, t_0\rangle$,
so that we can have a direct physical interpretation of the function $\phi(\vec x,t)$
as the time evolution of the  expectation value of  the field for that particular state.
Even though it is certainly not easy to find a solution to our problem 
for generic boundary conditions, we expect that under certain
circumstances it should be possible to determine $|\psi, t_0\rangle$.
In the following we consider a case, which is relevant to the physical
applications (see above), where $|\psi, t_0\rangle$ can be found.
Namely we take at $t=t_0$ a constant  field $\phi(\vec x,t_0)=\phi_0$ with vanishing 
derivatives up to the  $(2m)^{{\rm th}}$ order. Our goal is now to identify the state 
$|\psi ,t_0 \rangle$ defined through the variational principle. 

To this end we consider a different physical problem, namely the case 
in which, for $t > t_0$, the field 
$\phi$ is a constant : $\phi=\phi_0$. For this new problem and for 
$-\infty < t < t_0$, the source $J$ has to be 
equal to the one of the previous case ($J=\overline J({\vec x},t)$),
while for $t > t_0$, $J$ is given by the constant $J_0$, 
which corresponds to $\phi_0$. This constant is obviously fixed by the
condition: 
\begin{equation}
\label{geinot}
\left.{\frac{\delta \Gamma[\phi]}{ \delta \phi}}\right |_{\phi=\phi_0}
=J_0.
\end{equation}

With this assignment for the source $J$, 
the solution of Eq. (\ref{eqstd1}) for $t > t_0$ must be 
the time-independent state  
\begin{equation}
\label{stato}
|\psi  ( J_0) \rangle = |\psi ,t_0 \rangle.
\end{equation}
In fact in this case  Eq. (\ref{eqstd1}) becomes: 
\begin{equation}
\label{eqaut}
\left ( \hat H -J_0 \int {\rm d}^3 {\vec x} ~\hat \Phi({\vec x}) - E_0 \right ) |\psi  ( J_0) \rangle = 0
\end{equation}
where $w(t)=-E_0$ is now a time-independent constant, and  the two constraints
in Eqs. (\ref{c1}) and (\ref{c2}) are:
\begin{equation}
\label{newcon}
\langle \psi (J_0) |\hat  \Phi({\vec x}) | \psi (J_0) \rangle =\phi_0,~~~~~~~~~~~~~~~~~~~
\langle \psi (J_0)  | \psi (J_0) \rangle =1.
\end{equation}

Clearly this is nothing else than the time-independent variational 
principle that defines the effective
potential (see Eqs.(\ref{dep})--(\ref{eqsti})).  
We then have found that the state 
$|\psi ,t_0 \rangle=|\psi  ( J_0) \rangle$ we were looking for is
nothing but the normalized time-independent state  
that provides the minimum expectation value of the Hamiltonian $\hat H$
among the normalized states with field expectation value equal to $\phi_0$.

There is one main assumption behind this result, and  it concerns the 
source $J$. In fact, for our purposes, we have considered two
different sources,
which are both equal to $\overline J({\vec x},t)$  in the range 
$-\infty < t < t_0$, 
whereas for $t > t_0$ one is vanishing while the other is a non-vanishing constant $J=J_0$.
Note that, by construction, $J_0=\overline J({\vec x},t_0)$. Therefore in the former case 
the source is discontinuous at $t=t_0$  and we are assuming that this has no 
consequences on the determination of the solution of the time-dependent
Schr\"odinger equation. 
This discontinuity  could be replaced  with  a sharp but still continuous  
change of the  function $J$ around $t=t_0$, but again we would have to assume that this 
has no consequences on the determination of the state.

Finally, let us come back  to the important question, mentioned at the beginning 
of this section, that at least in some cases the effective action 
(or an approximation of it) can be obtained as a diagonal matrix element, 
i.e. in Eq. (\ref{pvea}) we have $|\psi_- ,t\rangle= | \psi_+ ,t\rangle $. 
We now argue that this should be the case when the motion of the field 
$\phi$ in the time interval $[t_0,t_1]$, i.e. the time interval where the motion of $\phi$
is governed by Eq. (\ref{qeq}) plus a set of boundary conditions, is periodic. 
Suppose that this time interval covers $n$ periods (let us call $\Delta t$ 
the period), i.e. that $t_1=t_0+n~\Delta t$. Then, at time
$t=t_1$, the field $\phi$ and its derivatives take the same 
values as at $t=t_0$. 

Now we can obviously make a shift on the time axis without changing the physics 
and take the point $t=0$ to be symmetric with respect to $t_0$ and $t_1$, i.e. 
$-t_0=t_1=\overline t$ in the new frame.   
Let us now choose $J$ in Eq.(\ref{eqstd1}) to be symmetric for time reflection,
namely $J$ must be taken in such a way that 
the ground state at $t=-\infty$ is driven 
onto $|\psi ,-\overline t \rangle$ at $t=-\overline t$ (and as before we indicate this 
particular source as $\overline J({\vec x},t)$ with $-\infty<t<-\overline t$ ), 
then $J=0$ within the range $[-\overline t, \overline t ]$ (in this interval $\phi$ covers 
$n$ periods), and, for  $\overline t<t<+\infty$,
the source  is turned on symmetrically, i.e.: 
$J(\vec x,t)=\overline J({\vec x},-t)$. 

Clearly, with such a choice,  the evolution of the two states in Eqs. (\ref{eqstd1}) 
and (\ref{eqstd2}) is symmetric for time reversal and therefore equal, i.e.  
$|\psi_+ ,t\rangle =|\psi_- ,t\rangle$. It must be noticed that the
class of periodic motions of $\phi$ obviously includes the case of constant 
$\phi$ (i.e. the case associated to a constant non-vanishing source), which 
was also considered above.

The conclusion of this section is that,
at least for those cases in which we consider a truncated derivative expansion 
of the effective action which gives rise to a periodic dynamical evolution of 
$\phi$, this latter quantity has the meaning of the expectation value of the field 
operator, and the associated quantum state, at some particular time
$t=t_0$,
can be identified with the state that defines the effective potential 
through the static variational principle (Eqs. (\ref{dep})--(\ref{eqsti})).

\section{An application. The harmonic oscillator}

We now discuss a simple application of the results  
presented in the previous section. As in \cite{jac}, we consider a 
trivial field theory (in one time 
and zero space dimension), namely the case of a one-dimensional harmonic oscillator 
whose Hamiltonian is ($\hbar=m=1)$:

\begin{equation}
\label{ho}
\hat H=\frac{\hat P^2}{2}+\frac{\omega^2 \hat
Q^2}{2}-\frac{\omega}{2}.
\end{equation}
The term $-\frac{\omega}{2}$ has been added in order to have a vanishing vacuum 
energy.

In this case it is known that the effective action $\Gamma$ 
coincides with the classical action $S$. In \cite{jac} this result has been recovered 
by means of the variational principle, which we illustrated in
Section 2, and the explicit solutions of Eqs. (\ref{eqstd1}) and (\ref{eqstd2}) 
for a generic source function $J(t)$ have been shown to be:

\begin{equation}
\label{hostat}
\langle Q|\psi_+ ,t\rangle =\langle Q|\psi_-,t\rangle=
\Bigl(\frac{\omega}{\pi}\Bigr)^{1/4}
~{\rm exp}\left \{ -\frac{\omega}{2}  (Q-q)^2-i \dot q (Q-q)-\frac{i
q\dot q}{2} \right \},
\end{equation}
with 
\begin{equation}
\label{hoq}
q(t)=\langle \psi_- ,t|\hat Q |\psi_+ ,t\rangle=
 \frac{i}{2} \int^{+\infty}_{-\infty} {\rm d}t' ~ J(t')~{\rm e}^{
 -i\omega|t-t'|}.
\end{equation}
Also, the Lagrange multiplier $w(t)$ turns out to be: $w(t)=J(t) q(t)$.
 
From Eq. (\ref{hostat}) we see   that $|\psi_+ ,t\rangle =|\psi_- ,t\rangle$ as 
was expected because of the periodicity of the motion of $q$ for the harmonic
oscillator effective action.
Moreover, we see that the right-hand sides 
of Eqs. (\ref{hostat}) and (\ref{hoq}) are totally determined once the source $J$ is given.

In addition we  show a concrete realization of the procedure
illustrated in section 3.
In fact, we derive from the effective action the equation of motion for $q(t)$ 
which is equal to the classical one (being in this case $S[q]=\Gamma[q]$) 
and we therefore have the typical solution 
\begin{equation}
\label{qsol}
q(t)=A~{\rm cos}(\omega t+\omega \tau),
\end{equation}
where the constant $\tau$ is fixed by the boundary conditions.
By taking  
\begin{equation}
\label{tau}
\tau=\frac{n\pi}{\omega} 
\end{equation}
with $n$ integer and $t_0=-\tau$ (where $t_0$ is defined in Section 3), 
we get $q(t_0)=A$ and $\dot q(t_0)=0$. 
Moreover $q(t)$ covers $n$ periods 
in the time range $\lbrack-\tau,\tau\rbrack$ 
and so $q(\tau)=A$ and $\dot q(\tau)=0$.

Our problem here is to show that the  source $J(t)$ that drives $q(t)$ from $q(-\infty)=0$ 
(in fact as seen in Section 3, the quantity $q$ at  $t=\pm \infty$ must be the expectation 
value of the coordinate for the ground state of the harmonic
oscillator, which is zero)
to the value $q(-\tau)=A$. Then, after the interval 
$-\tau < t < \tau$ where
$J=0$ and $q(t)$ has  the mentioned periodic behaviour, 
$q$ must evolve  from  $q(\tau)=A$ to $q(+\infty)=0$.

This source is 
\begin{equation}
\label{gei1}
J(t')=A~ \omega ~{\rm e}^{-\epsilon | t'| }~~~~~~~~~~~~~~~~~~~~~~~~~~
(-\infty < t' < -\tau)~{\rm and}~(\tau < t' < +\infty)
\end{equation}
\begin{equation}
\label{gei2}
J(t')=0
~~~~~~~~~~~~~~~~~~~~~~~~~~~~~~~~~~~~~~~~~~~~~~~~~~~~~~~~~~~~~~~~
(-\tau < t' < \tau)
\end{equation}
where $\epsilon$ is a small parameter,  which eventually has to be sent to zero.
In fact by inserting  this particular source into  Eq. (\ref{hoq}),
$q(t)$, for $-\tau < t < \tau$ is given by 
\bea
q(t)&=&\frac{i\omega A}{2}\lim_{\epsilon \to 0}~~\lim_{T\to+\infty} 
~\left\{ \int^{-\tau}_{-T} {\rm d}t' 
~{\rm e}^{-i\omega t}~{\rm e}^{i\omega t'}~{\rm e}^{\epsilon t'}+
\int_{\tau}^{T} {\rm d}t' 
~{\rm e}^{i\omega t}~{\rm e}^{-i\omega t'}~{\rm e}^{-\epsilon t'}
\right \}\nn \\
&=&A~{\rm cos}(\omega t+\omega \tau),
\label{qper}
\eea
where we have introduced the intermediate 
step of cutting the time integrals at a large time $T > 0$
 and then taking the limit $T\to +\infty$, which  will be helpful in the following calculations.
The final result in Eq. (\ref{qper}), which is the desired solution introduced in Eq. (\ref{qsol}),
 has been  obtained with the help of 
\begin{equation}
\label{iden}
{\rm exp}{(-i\omega \tau)}={\rm exp}{(i\omega \tau)},
\end{equation}
which  holds because of Eq. (\ref{tau}). 

We also check the behaviour of $q(t)$ for $t\to+ \infty$, as obtained from the source 
defined in Eqs. (\ref{gei1}) and (\ref{gei2}). 
We therefore compute the integral in Eq. (\ref{hoq}) by introducing as before the cutoff
$\pm T$ for large (positive and negative)  values of $t'$ and putting
$t=T$, and then taking the limits 
$T\to +\infty$ and $\epsilon\to 0$:
\bea
&&q(+\infty)=\frac{i\omega A}{2}\lim_{\epsilon \to 0}~~\lim_{T\to+\infty} ~\left\{ \int^{-\tau}_{-T} {\rm d}t' 
~{\rm e}^{-i\omega T}~{\rm e}^{i\omega t'}~{\rm e}^{\epsilon t'}+
\int_{\tau}^{T} {\rm d}t' 
~{\rm e}^{-i\omega T}~{\rm e}^{i\omega t'}~{\rm e}^{-\epsilon t'} \right \}
\nonumber\\
&&=\frac{i\omega A}{2}\lim_{\epsilon \to 0}~~\lim_{T\to+\infty} ~\left\{ 
\frac{  {\rm e}^{-i\omega T} }{i\omega+\epsilon}
~{\rm e}^{-i\omega \tau-\epsilon\tau}-
\frac{  {\rm e}^{-i\omega T} }{i\omega-\epsilon}
~{\rm e}^{i\omega \tau-\epsilon\tau}\right \}=0.
\label{qasy}
\eea

Instead of taking the limit $T\to \infty$ in the last line of Eq. (\ref{qasy}), 
we keep $T$ fixed to a very large value due to the oscillating terms ${\rm exp}{(-i\omega T)}$
and perform the limit $\epsilon \to 0$.
Then, by making use of Eq. (\ref{iden}), we find $q(+\infty)=0$.
The computation for $q(-\infty)$ is totally analogous and gives again $q(-\infty)=0$. 

Equations (\ref{qper}) and (\ref{qasy}) provide the desired results for the source 
$J(t)$ defined in Eqs. (\ref{gei1}) and (\ref{gei2}). As noted in
Section 3, there 
is a discontinuity in the function $J(t)$ at the two points $t=\pm
\tau$; however, in 
this particular example there is no consequence on the coordinate expectation 
value  (and its derivatives), which are continuous at those points.

According to Section 3  we must also find a new source that determines a 
static coordinate expectation value, namely $q(t)=A$  
for  $-\tau < t < \tau$. 
For this purpose, we retain the definition 
in Eq. (\ref{gei1})  for large positive and negative times but, 
following the argument in Section 3, instead of Eq. (\ref{gei2}),
we now take $J=A~ \omega ~{\rm exp}(-\epsilon \tau)$ constant  in the time interval $[-\tau, \tau]$
and this new source does not have any discontinuity points. 
Moreover in the two asymptotic regions the integrals for $t'$ are 
equal to the ones performed above and we have only to compute the contributions in the
region where the former source vanishes according to Eq. (\ref{gei2}).
Therefore, for $-\tau < t < \tau$, $q(t)$ is given by Eq.  (\ref{qper}) plus an additional term
(note that, when performing an integral  in this limited range $\lbrack -\tau,\tau\rbrack$  
the term ${\rm exp}(-\epsilon \tau)$ in the new source will always
tend to $1$ for  $\epsilon \to 0$ 
and it can therefore, for simplicity, be neglected from the
beginning):
\bea
&&q(t)=A~{\rm cos}(\omega t+\omega \tau)+\frac{i\omega A}{2}
~\int^{\tau}_{-\tau} {\rm d}t' 
~{\rm e}^{-i\omega |t-t'|}
\nonumber\\
&& 
=A~{\rm cos}(\omega t+\omega \tau)+\frac{A}{2} \left (2- {\rm e}^{-i\omega (t+\tau)}- {\rm e}^{i\omega (t+\tau)}
\right )=A,
\label{qlast1}
\eea
where again Eq. (\ref{iden})  has been used. This is exactly the required solution for $q$.

Finally, in order to check that $q(t)$ vanishes asymptotically even
with this source, we must  check that
the vanishing result in Eq.  (\ref{qasy})  is not modified. This time we have the additional contribution
\begin{equation}\label{qlast2}
q(+\infty)=\frac{i\omega A}{2}
~\int^{\tau}_{-\tau} {\rm d}t' 
~{\rm e}^{-i\omega (T-t')}
=\frac{A {\rm e}^{-i\omega T}}{2} \left (
 {\rm e}^{i\omega \tau}- {\rm e}^{-i\omega \tau} \right )
=0,
\end{equation}
which is vanishing because of Eq. (\ref{iden}).
The same is valid for $q(-\infty)$.

\section{The double-well potential}

As a further test of the arguments presented in Section 3, we now 
consider the non-trivial case of the motion of a wave packet in a 
double-well potential. To follow the dynamical evolution of the 
position operator expectation value, we make use of  Eq. (\ref{qeq}) 
within the framework of the LPA (see Eq.(\ref{lpa})) for the effective 
action. The effective potential $V_{eff}(q)$ is obtained with
the help of the wilsonian RG equation (see below), which, as we 
mentioned before, gives a non-perturbative, convex, effective potential 
(for comparison also the one-loop effective potential will be considered 
at the end of this section).
Again we fix for $q$ and $\dot q$ the two initial 
conditions: 

\be\label{incond}
q(t_0)=q_0 ~~~~~~~~~~~~~~~~~~~~~ {\rm and } ~~~~~~~~~~~~~~~~~~~~~ \dot q(t_0)=0. 
\ee
Then,
being again in the presence of a periodic motion, 
we are allowed to use Eq. (\ref{qeq}) as an equation of motion for  
$q(t)=\langle\hat Q\rangle_{_{t}}=\langle\psi,t|\hat Q|\psi,t\rangle$.

\begin{figure}[ht]
\begin{center}
\includegraphics[width=7cm,height=11cm,angle=270]{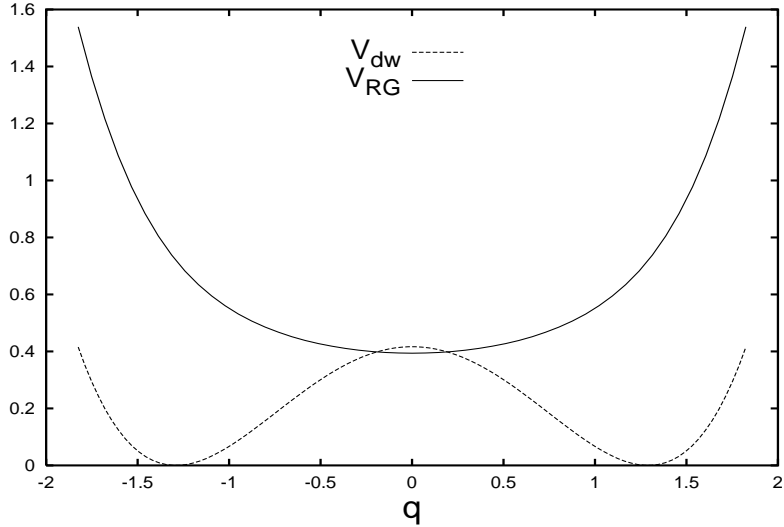}
\end{center}
\caption[]{$V_{dw}(q)$ and $V_{RG}(q)$ for $\lambda=0.15$.}
\label{fig1}
\end{figure}

According to our arguments, the function $q(t)$ derived by means 
of the above procedure should be equal to the one obtained by solving 
the time-dependent Schr\"{o}dinger equation with the form of the initial 
wave packet  fixed by Eq. (\ref{eqaut}). In fact we have numerically solved the 
latter problem with the initial conditions given by Eq. (\ref{eqaut});
for comparison, we also considered the evolution of an initial 
gaussian wave packet that is often used in the
applications\footnote{The RG flow equation for $V_{RG}$
and time-dependent Schr\"{o}dinger equation were
solved with the help of the NAG routines \cite{nag}. These routines
were also used to compute the mean values of $\hat Q$ and $\hat P$.}.

When we compare these results with 
the one  obtained within the framework of the effective action formalism, 
we see that, given the initial conditions (\ref{incond}), Eq. (\ref{qeq}) 
(together with Eq. (\ref{lpa})) provides a very accurate description of the 
dynamical evolution of $q(t)$, if the initial state 
is chosen according to our criterion, i.e. with the help of 
Eq. (\ref {eqaut}). We also see that the (non-convex) one-loop 
effective potential, that is the approximation to
$V_{eff}$ considered in the applications, is inadequate to describe 
the dynamics of $q(t)$. 

In Fig. \ref{fig1} we have plotted the double-well potential 
$V_{dw}(q)$, which we have written as: 
$V_{dw}(q)= -\frac{1}{2}q^2+\lambda q^4+\frac{1}{16\lambda}$, together
with the effective potential $V_{eff}(q)$, obtained by solving the
RG equation\cite{weg,hasen,twl,altog}:

\be\label{rge}
\frac{\partial U_k(q)}{\partial k} =
-\frac{1}{2\pi} {\rm ln} \left(1+\frac{U_k^{''}(q)}{k^2}  \right )
\ee

\nin
(here $U_k^{''}(q)$ means $\frac{\partial^2 U_k(q)}{\partial q^2}$).
Note that in this framework  the classical double-well potential 
$V_{dw}(q)$ is nothing but the bare potential, i.e. the UV 
boundary condition 
for $U_k(q)$; also note that the effective potential $V_{eff}(q)$ is 
approximated by the solution of Eq. (\ref{rge}), once it has been 
integrated down to $k=0$, i.e. once the 
quantum fluctuations have been taken into account: 
$V_{eff}(q) = V_{RG}(q) =  U_{k=0}(q)$.    
Also, when Eq. (\ref{lpa}) is inserted in Eq. (\ref{qeq}),  the 
quantum equation of motion for $q(t)$ takes the form:

\be\label{clas}
\ddot{q}=-\frac{d}{dq}\,V_{eff}(q) \, ,
\ee

\nin
i.e. the dynamical evolution of $q(t)$ is given by the classical 
equation of motion, being the classical potential replaced by $V_{eff}$, which, as 
discussed above, will be approximated by $V_{RG}$ in our numerical analysis. 

\begin{figure}
\begin{minipage}{5.5cm}
	\centerline{\includegraphics[width=7cm,height=6.5cm,angle=270]{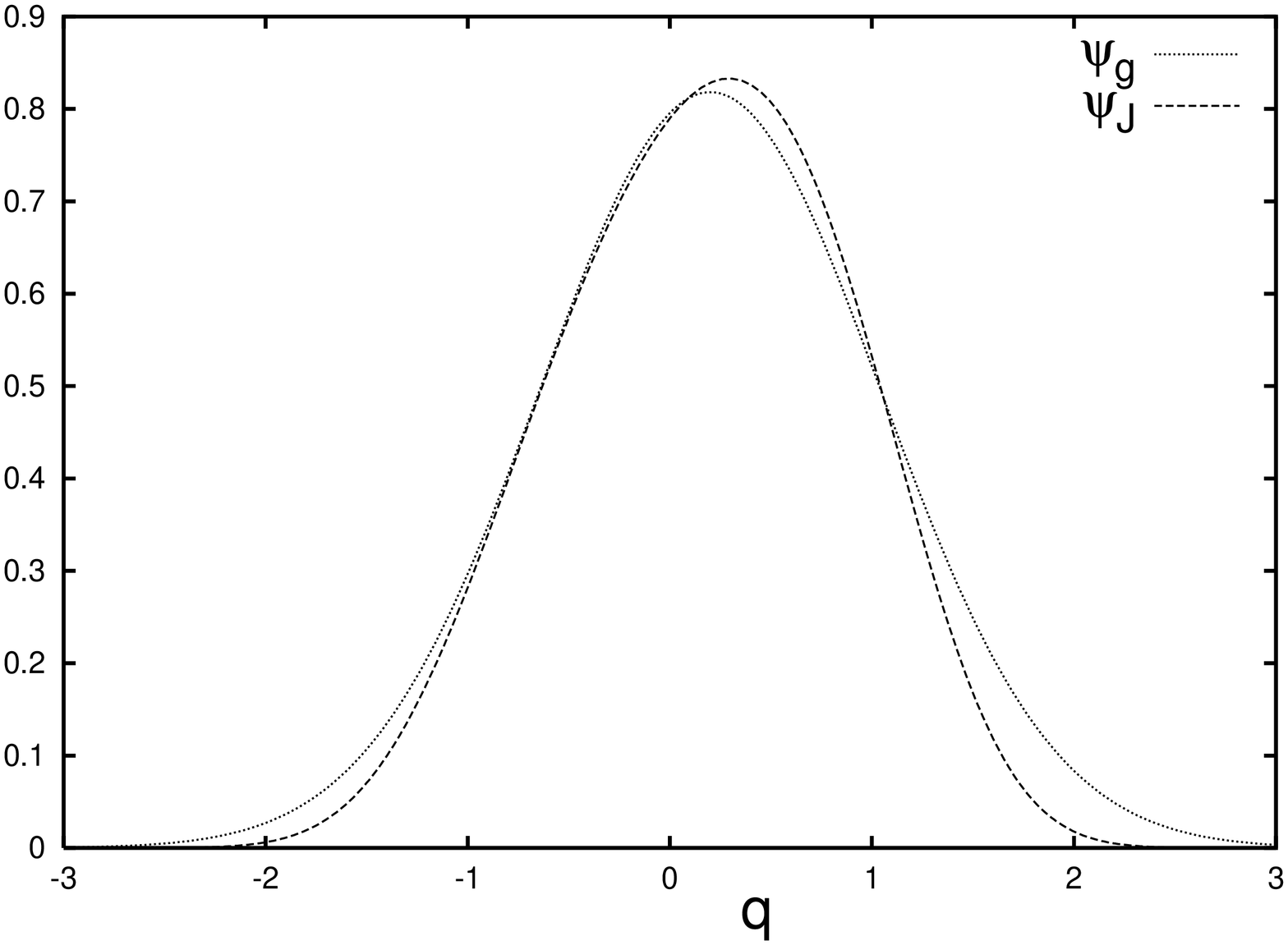}}
\centerline{(a)}
\end{minipage}
\hfill
\begin{minipage}{8.7cm}
	\centerline{\includegraphics[width=7cm,height=8cm,angle=270]{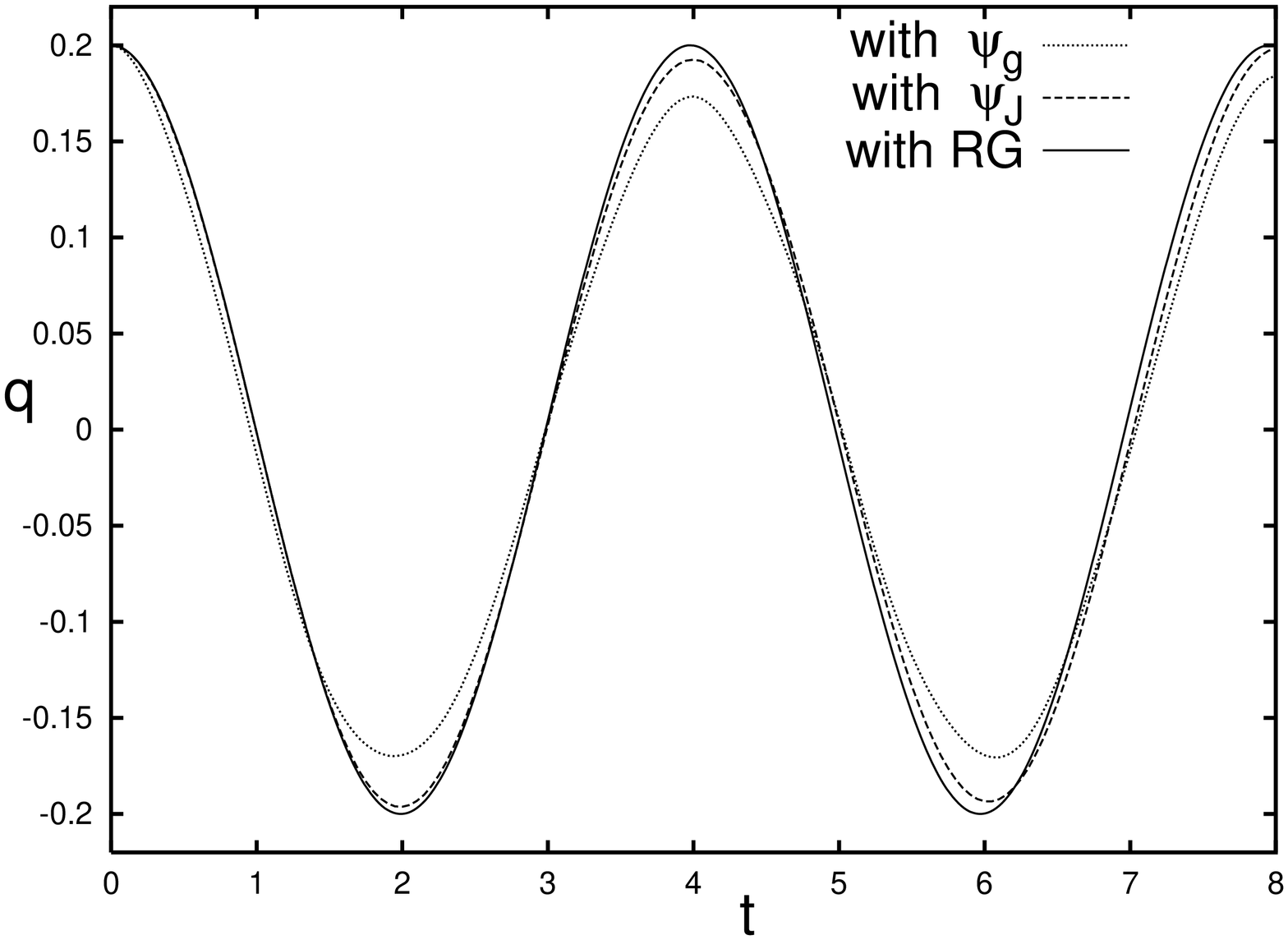}}
\centerline{(b)}
\end{minipage}
\caption{The case $\lambda=1.1$. (a). The initial wave packets $\psi_{_{J}}(q)$ 
and $\psi_{_{g}}(q)$. (b). The time evolution of $q(t)$ as obtained from
the Schr\"odinger equation with initial wave packets $\psi_{_{J}}(q)$ and 
$\psi_{_{g}}(q)$, and from Eq. (\ref{clas}) with the initial conditions
given in the text.}
\label{fig2}
\end{figure}

We now illustrate the above points by computing 
$q(t)$ and $p(t)~(=\dot q(t))$ for three different values of $\lambda$ : 
$1.1 \, , 0.15$ and $0.07$. We also choose the initial time as 
$t_0=0$, and the initial values  
of $q$ and $\dot q$ as follows: $q(0)=0.2$ for $\lambda = 1.1$,
$q(0)=0.5$ for $\lambda = 0.15$, 
$q(0)=1.0$ for $\lambda = 0.07$, and $\dot q(0)=0$ for each value of $\lambda$.

\begin{figure}[ht]
\begin{center}
\includegraphics[width=8cm,height=12cm,angle=270]{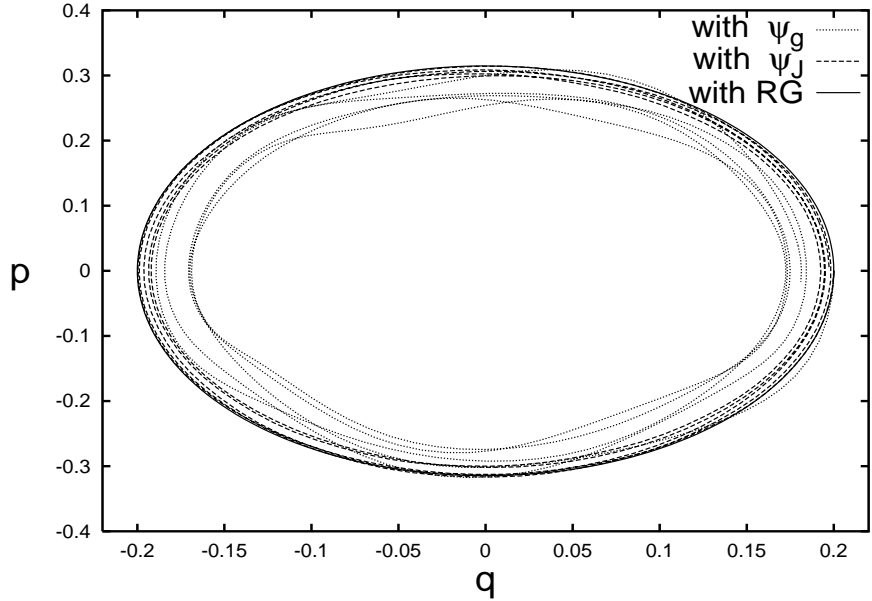}
\end{center}
\caption{The phase-space $q-p$ diagram for the three cases considered
in Fig. 2b ($\lambda=1.1$).}
\label{fig3}
\end{figure}

According to our arguments, for such a choice of the initial values, 
Eq. (\ref{qeq}) describes the time evolution of the expectation value $q(t)$ 
related to the initial wave packet, which is obtained by solving the 
time-independent Schr\"{o}dinger equation (see Eq. (\ref{eqaut})):

\be\label{scheq}
\left(-\frac{\hbar^2}{2m}\Delta+V_{dw}(q)-J\,q\right)\psi(q)=E_{_{J}}\,\psi(q)\, ,
\ee  

\nin
where, for each of the chosen initial values of $q(t)$, $J=J(q_{_0})$ 
is computed with the help of the equation: 
$\langle\psi_{_{J}}|\hat Q|\psi_{_{J}}\rangle= q_{_0}$ (for notational 
convenience we have replaced $q(0)$ by $q_{_0}$).  

Once the initial wave packet $\psi(q,0)$ is obtained with the help of 
Eq. (\ref{scheq}), $\psi(x,t)$ is obtained by solving the 
time-dependent Schr\"{o}dinger equation numerically, and 
the exact values of $q(t)$ and $p(t)$ are then computed. 
These results are compared with those obtained by solving 
Eq. (\ref{qeq}). 

As we already discussed in the Introduction, the boundary 
conditions that define the effective action are not appropriately taken 
into account in the applications 
of the effective action formalism to dynamical problems . 
Typically one associates a gaussian initial wave packet
with the initial conditions (\ref{incond}).
For this reason we have 
also considered the exact evolution of a gaussian wave packet,
$\psi_{_{g}}(q)$, chosen in such a way that its width is equal 
to that of  the ground state of the harmonic approximation at the bottom 
of one of the two wells. 

\begin{figure}
\begin{minipage}{5.5cm}
	\centerline{\includegraphics[width=7cm,height=6.5cm,angle=270]{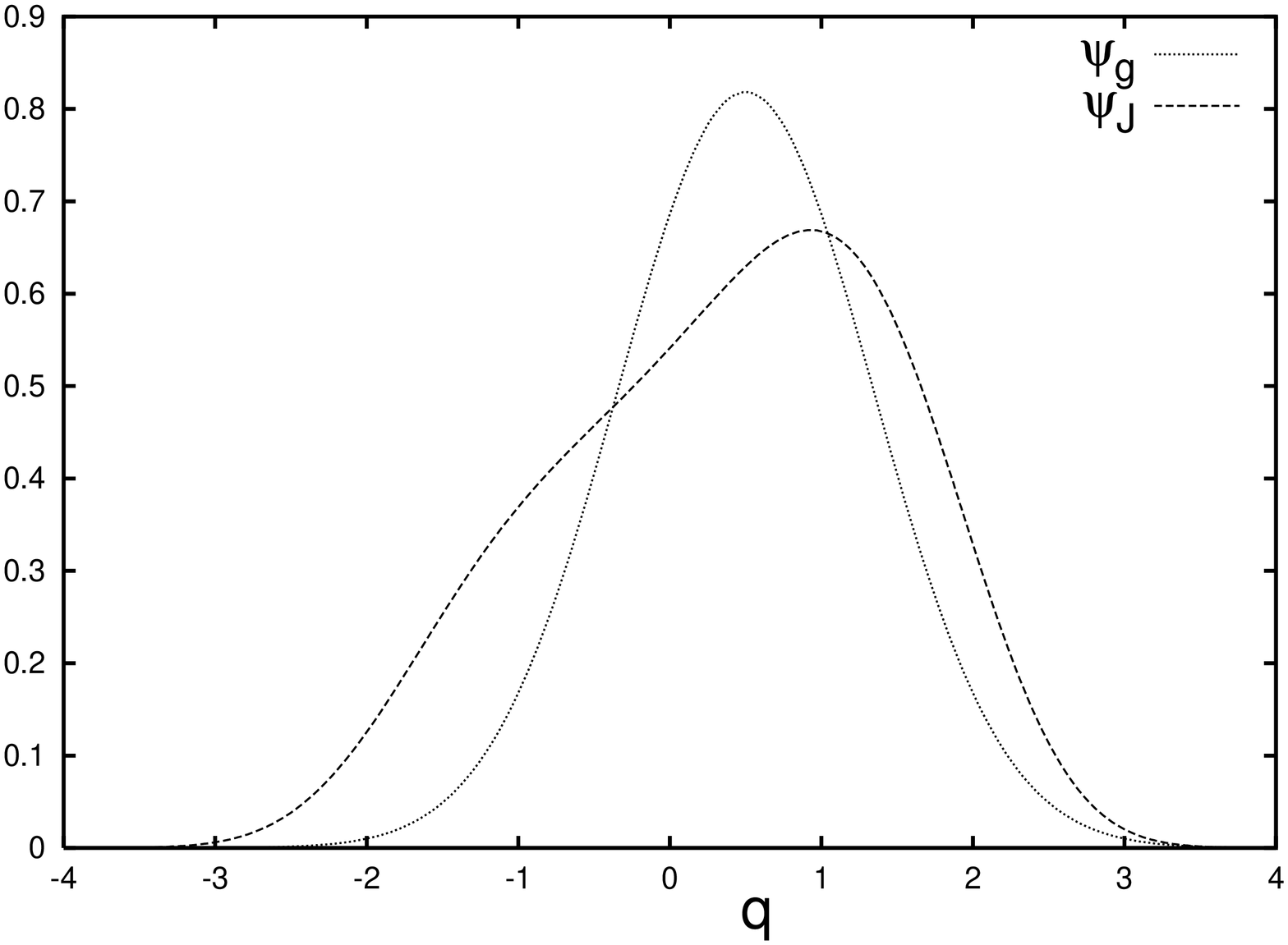}}
\centerline{(a)}
\end{minipage}
\hfill
\begin{minipage}{8.7cm}
	\centerline{\includegraphics[width=7cm,height=8cm,angle=270]{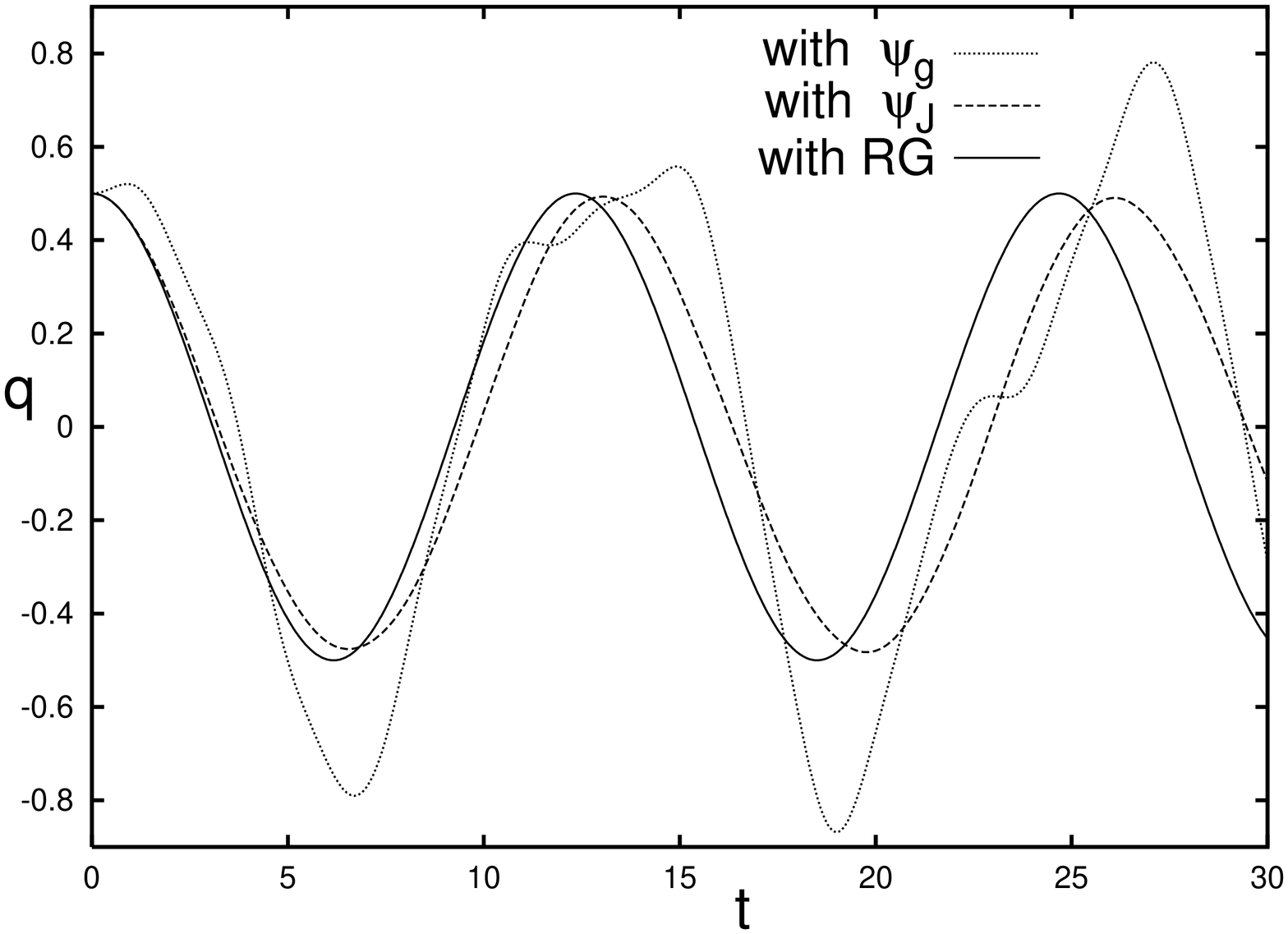}}
\centerline{(b)}
\end{minipage}
\caption{The case $\lambda=0.15$. (a). The initial wave packets 
$\psi_{_{J}}(q)$ and $\psi_{_{g}}(q)$. (b). The time evolution 
of $q(t)$ as obtained from
the Schr\"odinger equation with initial wave packets $\psi_{_{J}}(q)$ and 
$\psi_{_{g}}(q)$, and from Eq. (\ref{clas}), with the initial conditions
given in the text.}
\label{fig4}
\end{figure}

For each of the three values of $\lambda$ considered,
we present in the following figures the shape of the initial 
wave packets, $\psi_{_{J}}(q)$ and $\psi_{_{g}}(q)$, the 
expectation value of the position operator  
$\langle\hat Q\rangle_t=q(t)$ as a function
of time, and the phase-space, $q-p$, diagram. 

\begin{figure}[ht]
\begin{center}
\includegraphics[width=8cm,height=12cm,angle=270]{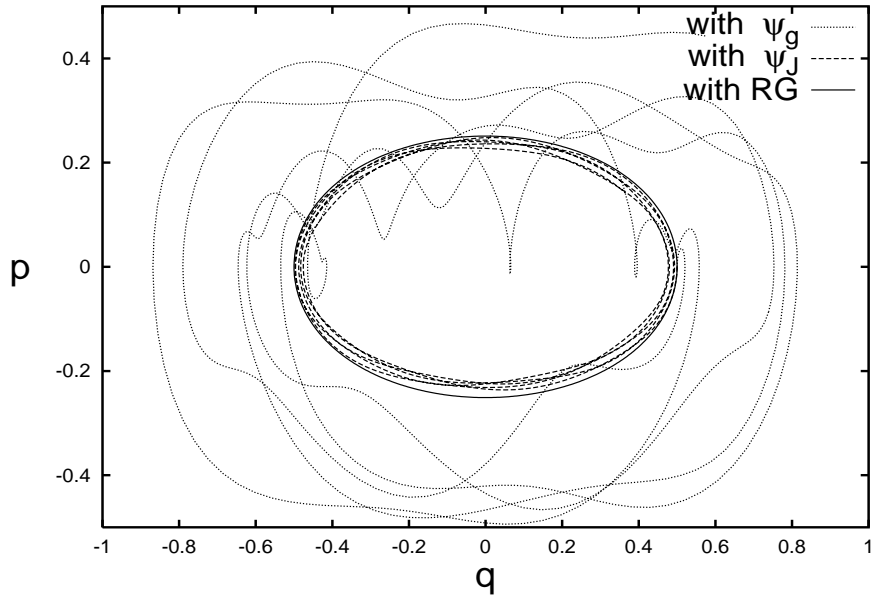}
\end{center}
\caption{The phase-space diagram for the three cases considered
in Fig. 4b ($\lambda=0.15$).}
\label{fig5}
\end{figure} 

In Figs. \ref{fig2} and \ref{fig3} we present the results 
for the $\lambda=1.1$ case. 
From Figs. \ref{fig2}b and \ref{fig3} we see that the results obtained with  
$\psi_{_{J}}(q)$ and $\psi_{_{g}}(q)$ are almost the same (and that
they are both well approximated by the results of Eq. (\ref{clas})). 
As is clear from  Fig. \ref{fig2}a, this is because, for 
this particular value of $\lambda$,  $\psi_{_{g}}(q)$ almost
coincides with $\psi_{_{J}}(q)$.  

However, we shall now see that, when we consider other values 
of $\lambda$, corresponding to different heights of the potential 
barrier, the choice of the correct initial wave packet becomes crucial. 
Taking for instance $\lambda =0.15$, we obtain two different shapes for 
$\psi_{_{J}}(q)$ and $\psi_{_{g}}(q)$ (Fig. \ref{fig4}a). It is not difficult
to imagine that they have a different dynamical evolution, as can 
actually be seen from Figs. \ref{fig4}b and \ref{fig5}. The crucial result 
for our analysis is that Eq. (\ref{clas}), i.e. the LPA of 
Eq. (\ref{qeq}), gives a very good approximation to the time 
evolution of $\langle\hat Q\rangle$ when the initial state is 
$\psi_{_{J}}(q)$ (see Fig. \ref{fig5}). This gives a very robust 
support to the arguments we have developed in Section 3.

\begin{figure}
\begin{minipage}{5.5cm}
	\centerline{\includegraphics[width=7cm,height=6.5cm,angle=270]{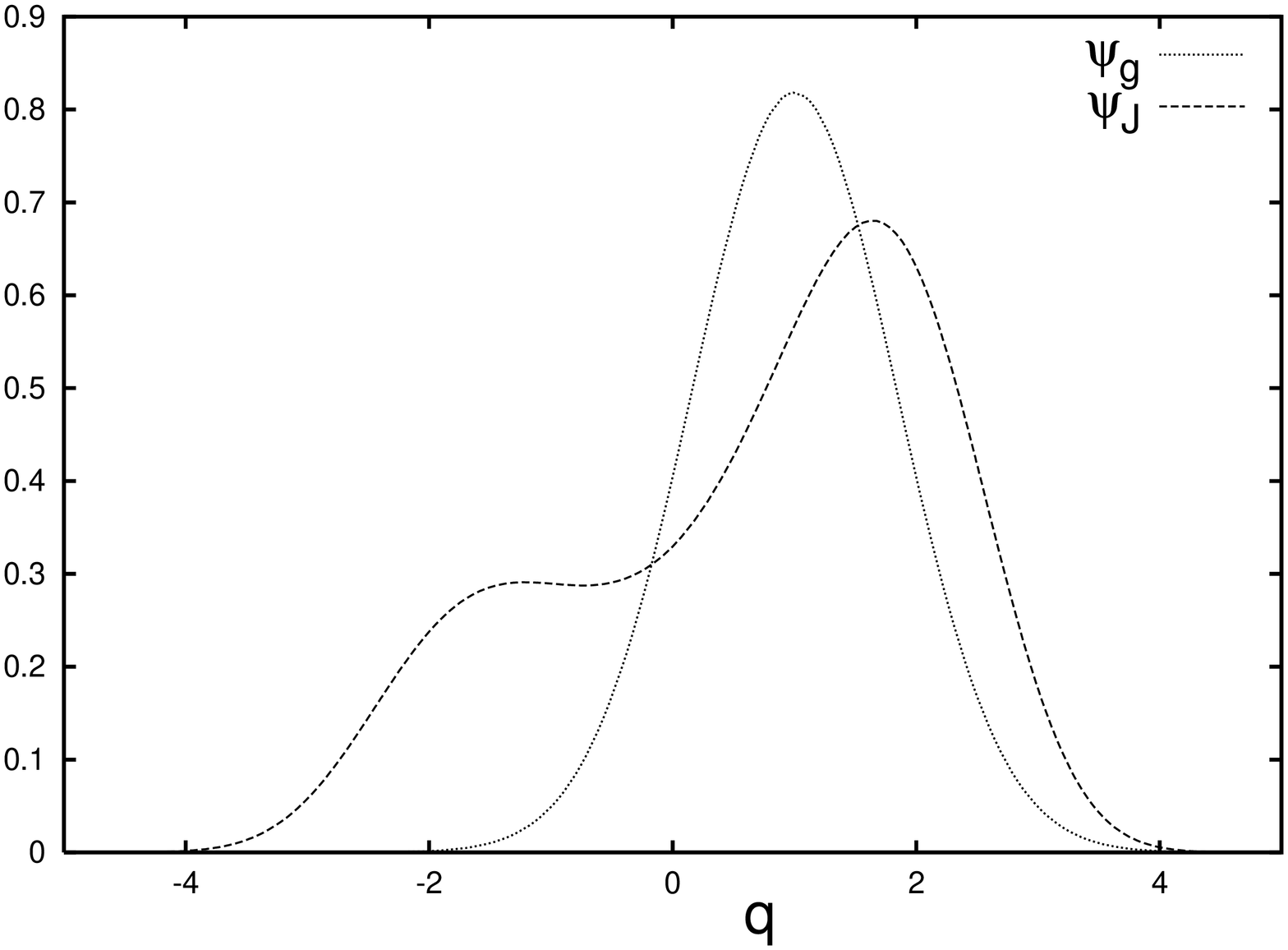}}
\centerline{(a)}
\end{minipage}
\hfill
\begin{minipage}{8.7cm}
	\centerline{\includegraphics[width=7cm,height=8cm,angle=270]{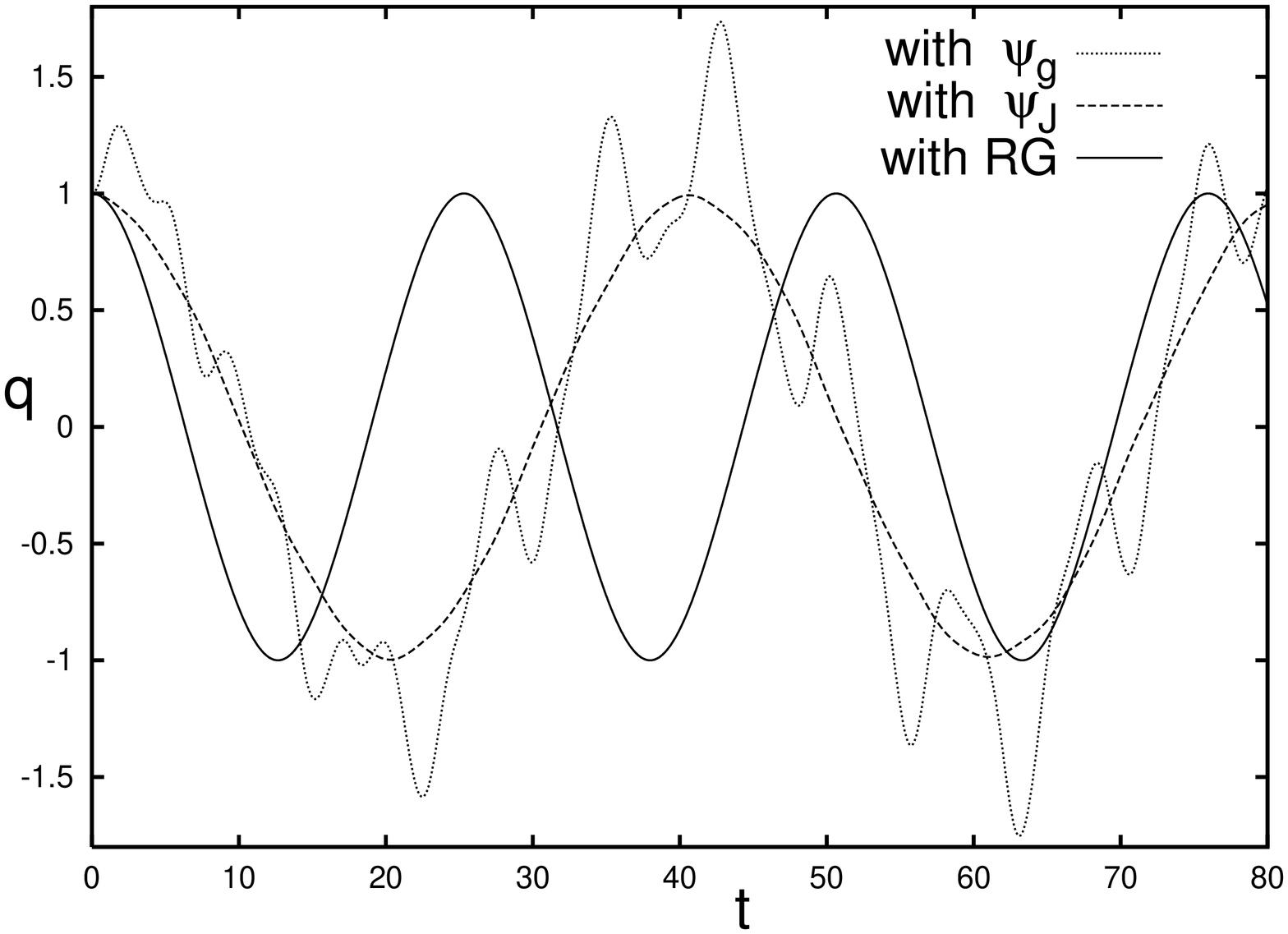}}
\centerline{(b)}
\end{minipage}
\caption{The case $\lambda=0.07$. (a). The initial wave packets 
$\psi_{_{J}}(q)$ and $\psi_{_{g}}(q)$. (b). The time evolution 
of $q(t)$ as obtained from
the Schr\"odinger equation with initial wave packets $\psi_{_{J}}(q)$ and 
$\psi_{_{g}}(q)$, and from Eq. (\ref{clas}) with the initial conditions
given in the text.}
\label{fig6}
\end{figure}

\begin{figure}[ht]
\begin{center}
\includegraphics[width=8cm,height=12cm,angle=270]{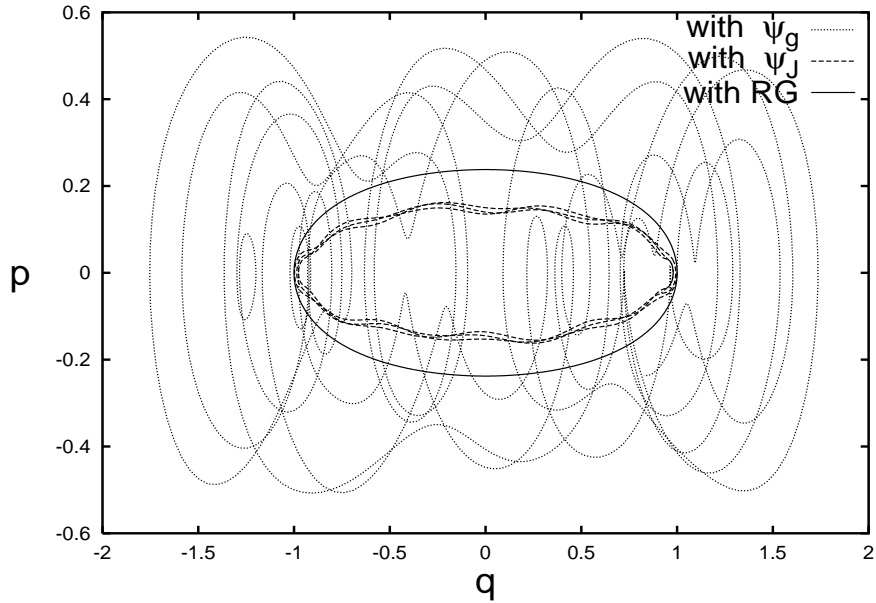}
\end{center}
\caption{The phase-space diagram for the three cases considered
in Fig. 6.b ($\lambda=0.07$).}
\label{fig7}
\end{figure}

As an additional example, we have considered the case  
$\lambda=0.07$. Once more, Figs.  \ref{fig6} and \ref{fig7}  
confirm that the effective
action formalism, in particular the LPA we have considered in 
the present paper, actually describes the dynamical evolution 
of the expectation value of the position operator once the 
initial state is selected according to Eq. (\ref{scheq}).

It is worth noting that, because of the initial conditions chosen for $\dot q$, 
the cases considered above describe tunnelling processes. Needless
to say, a semiclassical expansion of $\Ga$ at the lowest order, i.e. 
the approximation $\Ga=S$, would have been unable to describe these
tunnelling processes. In fact, in this approximation, 
$V_{eff}=V_{dw}$.  

\vfil
\eject

From the above results, it is clear that the trajectory of the gaussian wave 
deviates more and more  from the correct result  for decreasing values of $\lambda$,
i.e. for larger quantum effects due to the tunnelling.  Therefore it can be employed 
as an ansatz of the correct wave packet only for  a specific range of the coupling 
$\lambda$ and this is exactly what has been done in \cite{noiprl}, where the results 
obtained with a gaussian wave packet are still accurate (as is the
case for $\lambda=1.1$ as considered above).

This is a clear indication that a semiclassical approximation to 
Eq. (\ref{qeq}), which, for perturbatively small values of $\lambda$, would
correspond to the periodic motion of an almost gaussian wave packet around 
one of the classical minima, is misleading. The trajectory  obtained 
from Eq. (\ref{qeq}) includes quantum effects that are neglected in a 
semiclassical approach.

Figure \ref{fig7} needs one more comment. In fact for this smaller value 
of $\lambda$ the phase-space $q-p$ diagram obtained with the
``correct" initial wave packet, even though it is still
qualitatively  well described by the phase-space diagram obtained with
Eq. (\ref{clas}), shows a certain deviation from the latter. The reason for 
such a deviation has already been explained in \cite{noiprl} and has
to be traced back to the approximation that we have used for $\Ga$, 
the LPA. Actually, it was shown in \cite{noiprl} that the successive step in 
the derivative expansion of the effective action, i.e. the inclusion of
the wave-function renormalization in Eq.(\ref{lpa}), provides
an improvement of the LPA results. 
Here we are concerned with different problems, and so no longer 
pursue this issue.  

\begin{figure}
\begin{minipage}{5.5cm}
\centerline{\includegraphics[width=7cm,height=6.5cm,angle=270]{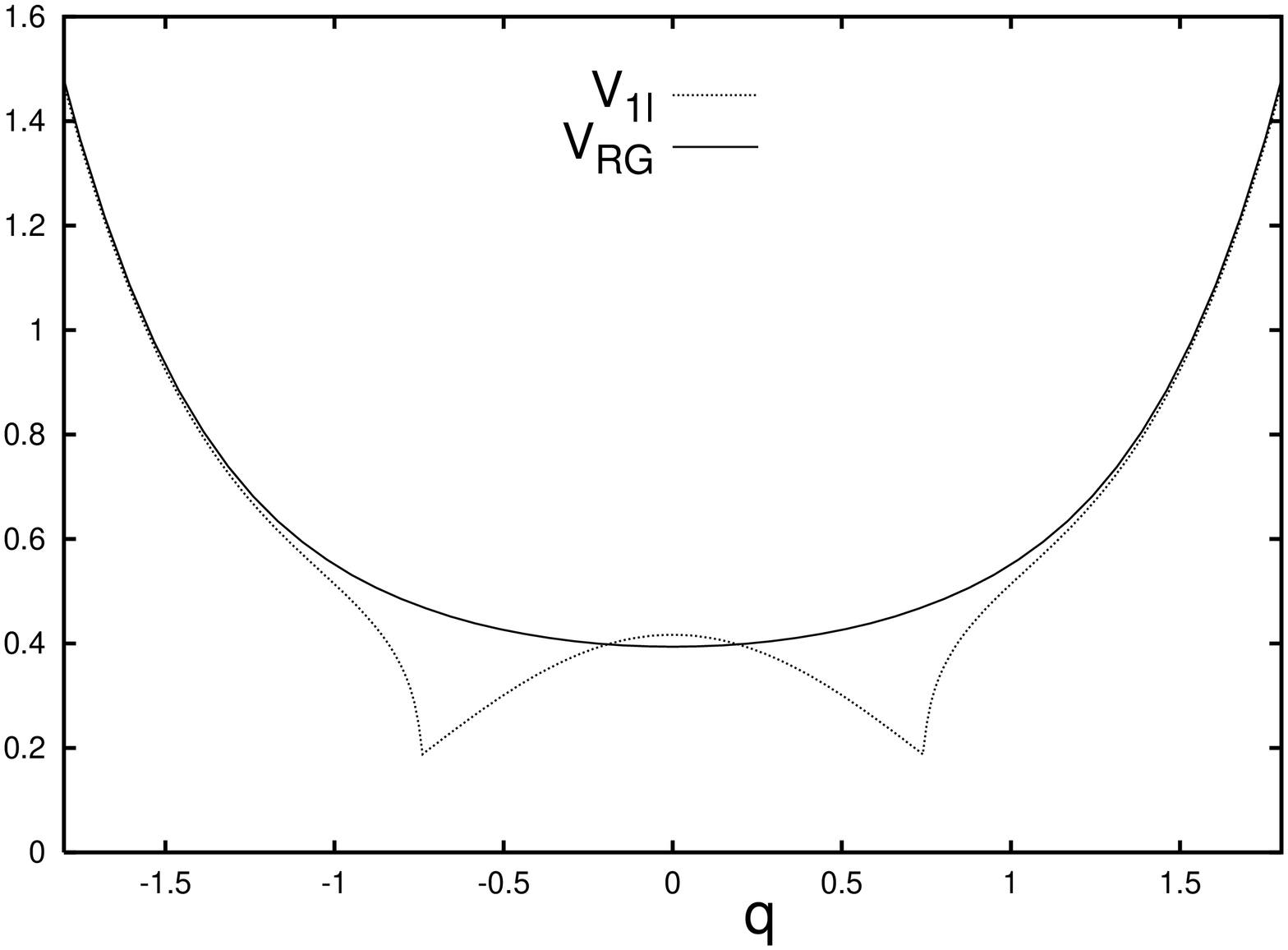}}
\centerline{(a)}
\end{minipage}
\hfill
\begin{minipage}{8.7cm}
	\centerline{\includegraphics[width=7cm,height=8cm,angle=270]{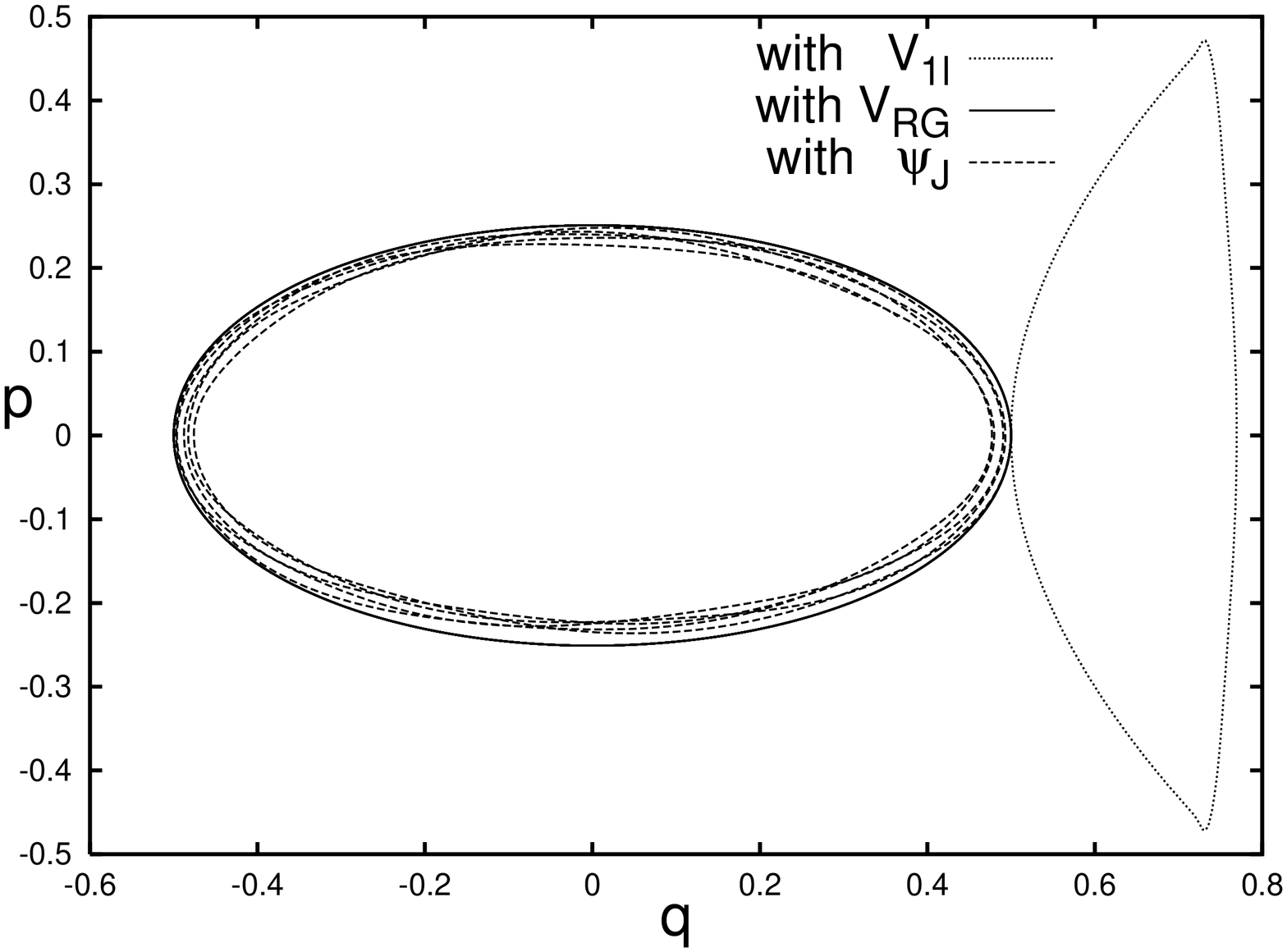}}
\centerline{(b)}
\end{minipage}
\caption{(a). The one-loop (non convex) effective potential 
compared with the RG (convex) one. (b). The phase-space diagrams
obtained from Eq. (\ref{clas}) with $V_{eff}$ approximated 
by $V_{1l}$ and $V_{RG}$, for a motion with 
$q(0)=0.5$ corresponding to an energy below the potential barrier.
Differently from $V_{RG}$, $V_{1l}$ completely fails in describing the
tunnelling. In (a) and (b) it is $\lambda=0.15$. }
\label{fig8}
\end{figure}

Before ending this section we would like to add a comment concerning
the approximation of $V_{eff}$,
namely the one-loop approximation: $V_{eff} \sim V_{1l}$. 
In Fig. \ref{fig8}a we show the one-loop effective potential versus
the RG effective potential. As is well known, the one-loop potential
does not enjoy the convexity property of the exact effective
potential $V_{eff}$. Actually this is true at any finite order of the
loop expansion for the effective potential. Moreover, in the region
between the two inflection points of the classical potential, the one-loop 
potential develops an imaginary part; according to the usual 
interpretation\cite{wewu}, this signals the instability of the
configurations in this region. In Fig.\ref{fig8}a the real
part of the potential is shown, which is the only part that is taken into
account in dynamical problems. 

\vfil
\eject

\begin{figure}[ht]
\begin{center}
\includegraphics[width=9cm,height=12cm,angle=270]{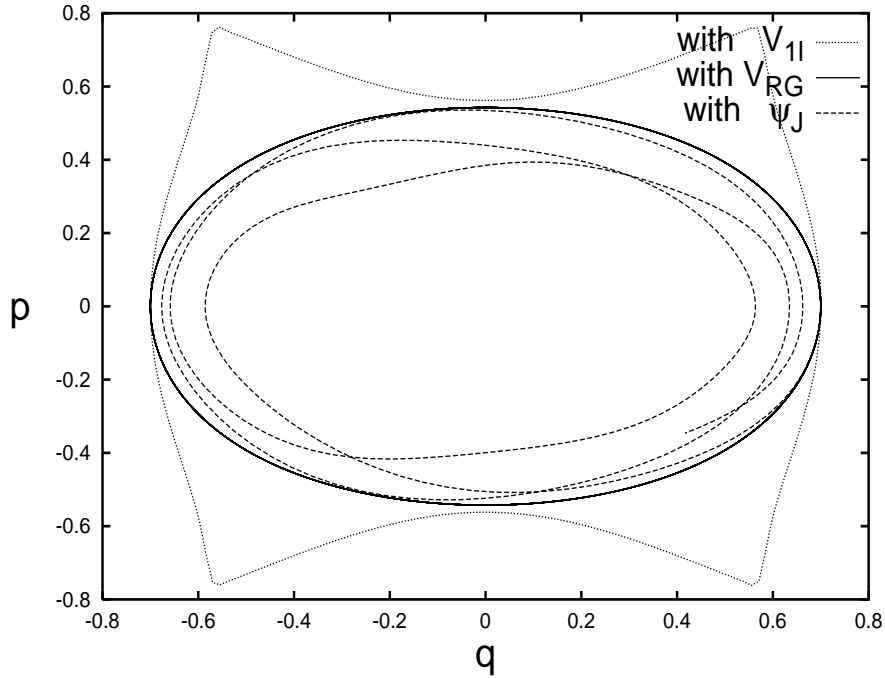}
\end{center}
\caption{The phase-space diagrams
obtained from Eq. (\ref{clas}), with $V_{eff}$ approximated respectively
by $V_{1l}$ and $V_{RG}$, for a motion with 
$q(0)=0.7$ corresponding to an energy above the potential barrier.
Even if now also $V_{1l}$ allows the whole $q$-region to be explored, again
we see that the motion is quite well described by $V_{RG}$ and not by
$V_{1l}$. Here we have taken $\lambda=0.25$.}
\label{fig9}
\end{figure} 

As is well known, $V_{1l}$ can be obtained analytically by replacing 
$U_k^{''}$ by $U_{\Lambda}^{''}$ in Eq. (\ref{rge}) 
(remember that $U_{\Lambda} = V_{dw}$), and then performing the elementary 
integration from $k=0$ to $k=\Lambda$, i.e. by solving the RG flow
equation in the independent-mode approximation. 
We thus obtain: $V_{1l} = V_{dw} + \delta V_{1l}$. Expanding in 
$\frac{1}{\Lambda}$ and neglecting $q$-independent terms, we can write
the one-loop correction $\delta V_{1l}$ as :

\begin{eqnarray}\label{loop}
\left \{ \begin{array}{ll}
\delta V_{1l}=\frac{1}{\pi} \sqrt{V_{dw}^{''}} \, \arctan \frac{\Lambda}
{\sqrt{V_{dw}^{''}}} + O(\frac{1}{\Lambda}) ~~~~~~~~~~~~~~~~~~~ \rm{for } 
|{\it x}|>\frac{1}{\sqrt{12\lambda}} \\
\delta V_{1l}=-\frac{1}{2\pi}\frac{V_{dw}^{''}}{\Lambda}
 + O(\frac{1}{\Lambda^2}) 
+ \frac{i}{2}\sqrt{-V_{dw}^{''}}
~~~~~~~~~~~~~~~~~~~ \rm{for } |{\it x}|<\frac{1}{\sqrt{12\lambda}}
\end{array} \right.
\end{eqnarray}
From the above equations we see that, in the region 
$|x|<\frac{1}{\sqrt{12\lambda}}$, ${\cal R}(\delta V_{1l})$ vanishes
with $\Lambda\to\infty$. A comparison of  Fig. \ref{fig8}a with Fig. \ref{fig1} 
also shows that, within this inner region, 
$V_{1l}$ is practically equal to $V_{dw}$. This is an example of the 
well-known fact that the loop expansion cannot change the concavity of $V_{dw}$.\\

It is clear that for the kind of
problems we have considered in this section, namely for tunnelling
processes, the one-loop effective potential $V_{1l}$ is as 
inadequate as the classical potential $V_{dw}$ to approximate  
$V_{eff}$ in Eq. (\ref{clas}). This is illustrated  in 
Fig. \ref{fig8}b, where it is shown that  for an initial
energy below the energy barrier the motion is obviously confined within one
well. According to the results of the semiclassical expansion, we can
actually state that at no finite order of this expansion can these 
phenomena be described by replacing the semiclassical effective
potential in Eq. (\ref{clas}). Instead they 
are perfectly well described by the solution of
Eq. (\ref{rge}), $V_{RG}$, which is only the
lowest-order approximation to $V_{eff}$ of a different expansion,
namely the gradient expansion. 

In order to push this comparison a step further, we have also
considered a motion with energy above the potential barrier.
Naturally,
in this case, by replacing $V_{1l}$ in Eq. (\ref{clas}) we obtain a
solution that, as is the case for $V_{RG}$, explores the whole
allowed region in position space. However, as is shown in
Fig. \ref{fig9}, the correct description of the motion is again given
by $V_{RG}$ and not by $V_{1l}$.

\section{Summary and outlook}

In the present work we have shown that in order to make use of the 
equation $\frac{\delta \Ga [\phi]} {\delta \phi(\vec x, t)} =0$ 
(Eq. (\ref{qeq}) in the text) as a ``quantum equation of motion",
attention has to be paid to a certain number of important issues. 

First of all we need to make sure that we are under conditions such
that the argument of $\Ga$, the classical field $\phi(\vec x,t)$,
is actually the expectation value of the quantum operator
$\hat\Phi(\vec x)$ in a given state $|\psi,t\rangle$, which is not 
always the case. 

Once we are under these conditions, we have shown that the asymptotic 
$|{\rm in }\rangle$ and $|{\rm out} \rangle$ states that enter the definition 
of the effective action have to be taken appropriately into account
in the determination of the initial wave functional $\Psi[\phi(\vec x,t_0)]$ 
associated with the motion of $\phi(\vec x,t)$.  
More precisely, by considering a derivative expansion for $\Ga$, 
 we have argued that it is possible to determine uniquely the initial state 
$\Psi[\phi(\vec x,t_0)]$ from the boundary conditions for the motion of $\phi(\vec x,t)$, 
assigned for instance on a given manifold $t=t_0$ (initial time), 
together with the asymptotic conditions encoded in the 
$|{\rm in }\rangle$ and $|{\rm out} \rangle$ states.
 
For initial conditions that are particularly important in the
applications, namely the case of constant initial $\phi$  
and vanishing derivatives of $\phi$ at $t=t_0$, we have been able to 
show that the initial (functional) wave packet obeys a modified 
time-independent {\it Schr\"odinger} equation, actually the equation that
allows the definition of the effective potential. 
 
Finally we have shown that, in the framework of the LPA for $\Ga$, where
the quantum equation of motion is obtained by the classical equation
of motion once the classical potential is replaced by the effective
potential, a reliable approximation to the effective potential comes 
from the solution of the corresponding wilsonian RG equation; instead,
the one-loop effective potential, and more generally any finite-order
approximation to the loop expansion for $V_{eff}$, fails to describe
the dynamics of $\phi(\vec x,t)$.

All the above points have been illustrated by considering quantum
mechanical examples. After a brief application of these results 
to the simple case of the harmonic oscillator, which confirmed our 
arguments, we have considered the case of a double-well potential.
The nice feature of considering a quantum-mechanical example is that 
we have the possibility to compute also the dynamical evolution of 
the wave packets with the time-dependent Schr\"odinger equation, 
i.e. we can compare our results with exact results. 
The cases we have investigated all confirm the correctness
of our conjectures. 

One point that was not considered in the present work concerns the
possible extension of these results to the case of higher
orders in the derivative expansion of $\Ga$.
It would also be interesting to investigate the impact of the presence 
of non-local terms in $\Ga$. We plan to investigate this point in future
work.

Naturally our ultimate goal is to apply our results in a QFT context
where, as we cannot solve the quantum mechanical (Heisenberg or
Schr\"odinger) equations exactly, we really need to have sensible
approximation schemes. We hope that our results can help in this
direction.

\vfill
\eject

\end{document}